\documentclass[12pt]{article}
\usepackage{bm}
\newcommand{\be}{\begin{equation}}

\newcommand{\ee}{\end{equation}}
\newcommand{\bea}{\begin{eqnarray}}
\newcommand{\eea}{\end{eqnarray}}

\newcommand{\de}{\partial}
\newcommand{\A}{\alpha}
\newcommand{\B}{\beta}

\newcommand{\eps}{\varepsilon}

\begin{document}
\hfill $ \vcenter{
  \hbox{\bf}
 \hbox{\bf } }$
\hfill{\bf BARI-TH 468/03 }\par \hfill{\bf CERN-TH/2003-171}\par
\hfill{\bf UGVA-DPT-2003-07/1107}
\begin{center}
{\Large\bf\boldmath {Quasi-particle Specific Heats  for the
Crystalline Color Superconducting Phase
of QCD }} \\
\end{center}
\begin{center}
{\Large\bf\boldmath {}} \rm \vskip1pc {\large R.
Casalbuoni$^{a}$\footnote{On leave from the Dipartimento di
Fisica, Universita' di  Firenze, I-50019 Firenze, Italia}, R.
Gatto$^b$, M. Mannarelli$^{c,d}$,
  G. Nardulli$^{c,d}$, M. Ruggieri$^{a,c,d}$ and S. Stramaglia$^{c,d}$} \\ \vspace{5mm} {\it{
$^a$TH-Division, CERN, CH-1211 Geneva 23, Switzerland
 \\
 $^b$D\'epart. de Physique Th\'eorique, Universit\'e de Gen\`eve,
 CH-1211 Gen\`eve 4, Suisse\\
  $^c$ Dipartimento di Fisica, Universit\`a di Bari, I-70124 Bari, Italia  \\$^d$
 I.N.F.N.,
 Sezione di Bari, I-70124 Bari, Italia\\
  }}
 \end{center}

 \begin{abstract}\noindent
We calculate the specific heats of quasi-particles of two-flavor
QCD in its crystalline phases for low temperature. We show that
for the different crystalline structures considered here there
are gapless modes contributing linearly in temperature to the
specific heat. We evaluate also the phonon contributions which
are   cubic in temperature. These features might be relevant for
compact stars with an inner shell in a color superconducting
crystalline phase.
\end{abstract}

\section{Introduction}
A number of theoretical studies have recently been devoted to QCD
at low temperatures $T$ and high densities. Besides the
theoretical interest for the different QCD phases, the possibility
of applications to compact stars, where dense quark matter might
exist, has driven much of the recent interest.

At high density and small $T$, quarks at the Fermi surface are
expected to condense, giving rise to color superconductivity, see
\cite{Barrois:1977xd,due} and, for reviews,
\cite{Rajagopal:2000wf}.
 At the largest densities
QCD with three flavors should be in the so called CFL
(color-flavor locked) phase. For decreasing density one expects
pairing between quarks of non vanishing total momentum, resulting
in crystalline gap behaviors
\cite{Alford:2000ze,sei,Giannakis:ab,Bowers:2002xr} (for reviews
see \cite{Casalbuoni:2003wh,nove}). Crystalline phases had
already been discussed for electric superconductors \cite{LO,FF},
so that one sometimes also refers to the crystalline phases as
LOFF phases. A crystalline color superconductor shell might exist
inside compact stars, between the external hadronic crust and the
internal CFL core.

In the present letter we discuss dispersion laws and specific
heats of the quasi-particles present in the LOFF phase. These
properties are relevant for the calculation of  the thermal
conductivity and neutrino emissivity and would affect the cooling
of the compact stars.

Thus far crystalline color superconductivity has been studied
only in a two-flavor model. Different crystal structures have
been computed, with the conclusion, based on a Ginzburg-Landau
discussion, that the most favored structure at zero temperature
is a cubic structure very close to a face centered cube (FCC).
However at non zero temperature the situation is far from being
clear (see the discussion in Ref. \cite{Casalbuoni:2003wh}),
therefore we will discuss  other crystalline structures as well.

The results that one can easily predict for the specific heat in
the LOFF phase with two flavors are a $T^3$ behavior for the
phonons and a $\mu^2 T$ behavior for the unpaired quarks. Our
subsequent analysis (Sections 3 and 4) confirms this expectation.
For the paired quarks one would naively expect an exponential
suppression $\exp(-\Delta/T)$, which is the behavior of the
regular 2SC phase. However our analysis shows (see Sect. 3) that
this is not the case, and for quark flavors that undergo LOFF
pairing the specific heat has a $\mu^2 T$ dependence. The reason
for this behavior is in the fermi quasi-particle dispersion law
(see Sect. 2) that for  the LOFF phase is  gapless. As a
conclusion, in a neutron star, provided a LOFF phase is present,
the specific heat and the thermal conductivity should be dominated
by fermions, with paired and unpaired quarks on the same footing.
A similar dominance is expected also for the LOFF phase with three
flavors. In this case there are no unpaired quarks and therefore a
a clear distinction should emerge in the thermal properties
(specific heat, thermal conductivity) of  the homogeneous case
(CFL) and the inhomogeneous one (LOFF) because all the quarks in
the former case would have the exponential suppression, while in
the latter the $T$ dependence is expected to be linear.

 The plan of the paper is as follows. In section 2 we discuss the
dispersion laws of the fermionic quasi-particles for different
crystal structures of the LOFF phase of high density QCD. We
discuss the one-plane wave structure, the strip, and the face
centered cube. In section 3 we calculate the contribution of the
fermionic quasi-particles to the specific heat for the three
mentioned structures. In section 4 we use the effective phonon
Lagrangian to calculate the phonon specific heats. Section 5 is
devoted to the conclusions.

\section{Fermi quasi-particle dispersion law}
In this section we  derive the Fermi quasi-particle dispersion
law in the two-flavor QCD LOFF phase for a few different
crystalline structures. We will work  with an inhomogeneous
condensate given by
\begin{equation}
\hat{\Delta}({\bf r})=\Delta({\bf r})\,\epsilon_{\alpha\beta
3}\,\epsilon_{ij} \label{condensates}
\end{equation}
where $\alpha,\beta$ are color indices and $i,j=1,2$ are flavor
indices. Notice that the gap term pairs together the color 1 and
2, whereas the fermions with color 3 are unpaired. We work in
presence of a difference in the chemical potential of the $u$ and
the $d$ quarks. We define
\begin{equation}
\mu_u=\mu+\delta\mu~,~~~~~~~~~\mu_d=\mu-\delta\mu~.
\end{equation}
As usual the quasi-particle dispersion law is obtained by looking
at the zeros of the inverse propagator. For fermionic
quasi-particles one has  to solve the eigenvalue equation
 \be (S^{-1})_{\A\B}^{ij} \chi_{j}^{\B}=0 \label{eig1} \, ,\ee
 where $S^{-1}$ is the inverse propagator in the LOFF phase
 in the Nambu-Gorkov
 formalism and $\chi_{j}^{\B}$ are the Green eigenfunctions.
  Following the notations in \cite{Casalbuoni:2003wh} we
 write $S^{-1}$ as follows:
 \be (S^{-1})_{\A\B}^{ij}=
\left(
\begin{array}{cc} \delta_{\A\B} [  \delta_{ij} (E + i {\bf v}
\cdot {\bm \nabla}) + \delta\mu(\sigma_3)_{ij}] & -\eps_{\A\B
3}\eps_{ij}\Delta({\bf r}) \cr -\eps_{\A\B 3}\eps_{ij}\Delta({\bf
r})^{*} & \delta_{\A\B} [ \delta_{ij} (E - i {\bf v} \cdot {\bm
\nabla}) + \delta\mu(\sigma_3)_{ij}]
\end{array} \right)\ee
where $E$ is the quasi-particle energy and $\bf v$ is the Fermi
velocity, that in QCD with massless quarks satisfies $v=|{\bf
v}|=1$. Let us define \be\chi_i^\alpha=\left(\matrix{\bar
G_i^\alpha \cr -i(\sigma_2)_{\alpha\beta}\bar F_i^\beta}\right)\,
. \ee Performing the unitary transformation \be \bar
G_i^\alpha=\left(e^{i\delta\mu\, \sigma_3\, {\bf v\cdot
r}/v^2}\right)_{ij} G_j^\alpha\,,~~~\bar
F_i^\alpha=\left(e^{-i\delta\mu\, \sigma_3\,{\bf v\cdot
r}/v^2}\,\sigma_2\right)_{ij} F_j^\alpha\ ~,
\label{transformation}\ee one measures the energy of each flavor
from its Fermi energy. The resulting equations for $F_i^\alpha$
and $G_i^\alpha$  are independent of color and flavor indices,
that therefore will be omitted below: \bea (E + i {\bf v}\cdot
{\bm\nabla}) G - i \Delta({\bf r})  F
 &=& 0\ ,
\cr (E  - i {\bf v}\cdot {\bm\nabla}) F+ i \Delta({\bf r})^{*}  G
&=& 0\ .\label{eigfg}
 \eea
We solve these equations for three different crystalline
structures, corresponding to different decompositions of
(\ref{condensates}) in plane waves: \begin{enumerate}
    \item One
plane wave;
    \item Two antipodal plane waves;
    \item The face centered cubic (FCC) structure, formed by eight
    plane waves with momenta pointing to the vertices of a cube.
\end{enumerate}
The first case is discussed for metals in \cite{FF} by Fulde and
Ferrel (FF); for QCD is analyzed in \cite{Alford:2000ze}. The
second case has been examined at $T=0$ by \cite{LO}; we will refer
to it as the {\it strip} below. The last case is the preferred
solution at $T=0$ in Ginzburg-Landau analysis of Ref.
\cite{Bowers:2002xr}.  The reason why we examine these different
structures is in the fact that the preferred structure at $T\neq
0$ can be different from that at $T=0$, see e.g
\cite{Giannakis:ab} and the discussion in
\cite{Casalbuoni:2003wh}.
\subsection{One plane wave \label{1wS}}

 For the FF condensate we take the direction of the
 Cooper pair total momentum
 $2\bf q$ along the $z-$axis. The gap
     \be \Delta({\bf r})=\Delta e^{2i q z} \ee
is therefore a complex number. If we take $ G =\hat{G}({\bf
r})e^{i{\bf q}\cdot {\bf r}}$, $ F =\hat{F}({\bf r})e^{-i{\bf q}
\cdot {\bf r}}$, we get from Eqs.(\ref{eigfg}) \bea  \left[E - q
v_z +i{\bf v \cdot \bm\nabla} \right] \hat{G}({\bf r})  & = &+\, i
\Delta \hat{F}({\bf r})\ ,\cr \left[E  -   q v_z -i {\bf  v \cdot
\bm\nabla}\right] \hat{F}({\bf r}) & = & -\,i \Delta \hat{G}({\bf
r})
  \label{eig1W}\ . \eea
These are the standard Gorkov equations for a uniform
superconductor with energy $E- qv_z$. The eigenfunctions are
simple plane waves \be \hat{G}({\bf  r})= u e^{i\bf k \cdot r}
\hspace{1.5cm} \hat{F}({\bf  r})=we^{i\bf k \cdot  r} \, ,
\label{bloch3} \ee and the quasi-particle spectrum is given by:
 \be E_\pm = qv_z \pm
\sqrt{\xi^2+ \Delta^2} \, ,\label{result}\ee where $\xi=\bf k\cdot
v$ is the residual longitudinal momentum, i.e. the longitudinal
momentum measured from the Fermi surface. Because of the
transformation (\ref{transformation}), quasi-particle energies are
computed from the corresponding Fermi energies $\mu_{u,d}$.
Eq.(\ref{result}) is the dispersion law of quasi-particle
($E_\pm\ge 0$) or hole states ($E_\pm<0$). We notice that in this
case, as in \cite{Alford:1999xc}, \cite{Shovkovy:2003uu},
\cite{Huang:2003xd} and \cite{Gubankova:2003uj} , we are in
presence of gapless superconductivity.

An anisotropic dispersion law was also obtained in
\cite{Alford:2000ze} by a different procedure (variational
method). Their result was obtained for finite $\mu$ and reduces to
(\ref{result}) if one considers only the leading order in the
asymptotic $\mu\to\infty$ limit, which is the limit we consider in
the present paper.
\subsection{Strip  and FCC structures}
Both the strip and the cubic structure have real $\Delta({\bf
r})$. The solutions of Eqs. (\ref{eigfg}) are Bloch functions \be
G = u({\bf r}) e^{i \bf k \cdot r}\ , \hspace{1.5cm} F = w({\bf r
}) e^{i
 \bf k \cdot r} \, , \label{bloch}\ee with $u({\bf r} )$ and
$w({\bf r})$ periodic functions and $\bf k$ in the first Brillouin
zone. They satisfy \bea \left[E - \xi +i {\bf v \cdot\bm \nabla}
\right] u({\bf r})  & = & +i\, \Delta({\bf r}) w({\bf r}) \ ,\cr
\left[E  +  \xi -i{\bf v \cdot \bm\nabla} \right] w({\bf r})  & =
& - i \,\Delta({\bf r}) u({\bf r})
 \label{eig3}\ . \eea
The corresponding  quasi-particles are gapless \cite{LO}. In fact,
for $E=0$ and $ \xi=0$, the system (\ref{eig3}) has two solutions.
We have $w_\pm=\pm u_\pm$, with $u_\pm$ solutions of \be {\bf v
\cdot \bm \nabla} u_\pm({\bf r}) = \pm \Delta({\bf r}) u_\pm({\bf
r})\, ,\ee and given by  \be u_{\pm}({\bf r}) =  \exp \Big[ \pm
\int \Delta({\bf r}^{\prime} ) \frac{d ({\bf r}^{\prime} \cdot
{\bf v})}{v^2} \Big]\ .\label{real}\ee Here the integration is
over a path joining the origin to the point ${\bf r}$.

 To find the dispersion law of quasi-particles
for small values of $\xi$ one uses degenerate perturbation theory
and gets \be E^2= \frac{\xi^2}{ A_+A_-} \, , \label{eps1}\ee with
\be A_\pm = \frac{1}{V_c} \int_{\mathrm{cell}} d \vec r
\exp\left[{\pm 2 \int \Delta({\bf r}^{\,\prime})\frac{d ({\bf
r}^{\,\prime} \cdot {\bf v})}{v^2} }\right] \, , \label{DEFA}\ee
where $V_c$ is the volume of a unit cell of the lattice.

Let us now specialize to the case of the strip, i.e. the
crystalline structure formed by two plane waves with wave vectors
$\pm 2\bf q$. The inhomogeneous gap is given by
\begin{equation}
\Delta({\bf r})=\Delta\,\cos\,2qz\  \label{stripGap}\end{equation}
and one gets
\begin{equation}
A_\pm^{(\mathrm{s})}\equiv A^{(\mathrm{s})}
=I_0\left(\frac{\Delta}{q v \cos\theta} \right) \,
,\label{Astrip}
\end{equation}
where $I_0(z)$ is the modified Bessel function of the zeroth
order.

Let us now turn to the FCC structure. Summing the eight plane
waves, the condensate can be put in the form
\begin{equation}
\Delta({\bf{r}})= \Delta\, \cos\,2qx\,\cos\,2qy\,\cos\,2qz \, .
\label{FCCGap}
\end{equation}If we define
 \bea B&=& \frac v 4\left(
\frac{\sin\,2q(x+y+z)}{v_x+v_y+v_z}+\frac{\sin\,2q(x+y-z)}{v_x+v_y-v_z}+
\frac{\sin\,2q(x-y+z)}{v_x-v_y+v_z}\right.
\cr&+&\left.\frac{\sin\,2q(-x+y+z)}{-v_x+v_y+v_z} \right) ,
\label{defF} \eea   one obtains for the cube\begin{equation}
A^{(\mathrm{fcc})}_\pm=\left(\frac{q}{\pi}\right)^3
\int_{cell}dV\, \exp\left\{\pm\frac{\Delta}{q v}B\right\} \,
,\label{AformulaCUBE}
\end{equation}where the integration is over the elementary cell of
volume $(\pi/q)^3$.

Let us  notice that whereas for the strip in Eq. (\ref{Astrip})
one has a closed formula, for the FCC  only numerical or
approximate expressions can be given. Since the parameter
$x={\Delta}/{q v}$ is expected to be small,  expanding
 $A^{(\mathrm{fcc})}_\pm$ as a power series one obtains \be
\sqrt{A^{(\mathrm{fcc})}_+(x)A^{(\mathrm{fcc})}_-(x)}= 1+  {\cal
O}[x]^2 \,
 . \label{dev}\ee
In conclusion for both the strip and the FCC the quasi-fermion
spectrum is gapless. This result remains valid also for massive
quarks, since the effect of the quark mass can be accounted for by
reducing the quark velocity from the value $v=1$ valid for the
massless case \cite{Casalbuoni:2002st}.

\section{Specific heat of the Fermi quasi-particles} \noindent
The contribution of the Fermi quasi-particles to the specific heat
per unit volume is \be c_v = \,\rho\, \int \frac{d \Omega}{4 \pi}
\, \int d \xi \, E \, \frac{d n(E,T) }{d T} \, ,
\label{generalCformula}\ee where, for the two flavor case (and
$|{\bf v}|=1$), \be\rho=\frac{4\mu^2}{\pi^2}\ ,\label{25}\ee while
$n(E,T)$ is the Fermi distribution function and the angular
integration is over the directions of $\bf v$.

We should also remind that, in the case of two flavors, the two
quarks of color 3 are unpaired and as such each contributes to the
specific heat as follows:\be c_v=\frac{\mu^2}3\,T\,.\label{26}\ee
Let us stress however that quarks that do not participate in the
standard spin 0 condensate can nevertheless couple with spin 1,
with much smaller gaps. Examples of these models can be found in
\cite{sba}.

 Let us now
specialize Eq. (\ref{generalCformula}) to the three crystalline
structure under scrutiny. We  limit the analysis to the small $T$
range.
\subsection{One plane wave}
The dispersion law  of quasi-particles  is given by
(\ref{result}). Using (\ref{generalCformula})  one gets in the
small temperature limit ($T \ll \Delta$) and for $\Delta < q $
 \be
c_v^{(\mathrm{FF})} = \frac{\rho T\pi^2 }3 \sqrt{1-
\frac{\Delta^2}{q^2}}\,~~~~~{\rm (quarks)}\ \, . \label{C1w}\ee
The specific heat depends linearly on  temperature because the
quasi-particle dispersion law (\ref{result}) gives rise to gapless
modes. There is also a contribution to the specific heat that
comes from gapped modes, but this contribution is exponentially
suppressed with the temperature.

\subsection{Strip  and FCC structures}
From Eq. (\ref{generalCformula}), using (\ref{eps1}) that is valid
for the strip and the FCC alike, one obtains, for $T \ll \Delta$:
 \be c_v =\, \frac{\rho T \pi^2}{3}\, \int
\frac{d \Omega}{4 \pi} \sqrt{A_+A_-}\ .\label{CA}\ee This
expression can be evaluated in closed form for the strip, when
$A_\pm=A^{(\mathrm{s})}$. One gets \be
c_v^{(\mathrm{s})}=\frac{\rho T\pi^2}{3}~
_1F_2\left(-{1}/{2};\,{1}/{2},\,1;\,\left({\Delta}/(q v
)\right)^2\right) \, ~~~~~~~~~~~~{\rm (quarks)}\label{Cstrip}
\end{equation}
where $_1 F_2$ denotes the generalized hypergeometric function
\cite{grads}. Differently from the analysis of \cite{LO}, here
$v$ is not small and  we can take $z= \Delta/q v\to 0$ near the
second order phase transition. Since for small $z$ one has $_1
F_2(-1/2;\,1/2,\,1; z^2)\simeq 1-z^2$, it is easily seen that the
normal Fermi liquid result (\ref{26}) is obtained for $z=0$. On
the other hand, at finite $\Delta$, the specific heat turns out
to be smaller.

For the face centered cube, $A^{(\mathrm{fcc})}_\pm$ can be
computed by evaluating the power expansion. One gets for this case
\be c_v^{(\mathrm{fcc})} = \frac{\displaystyle\rho
T\pi^2}{\displaystyle 3}\displaystyle \int\frac{\displaystyle
d\Omega}{\displaystyle 4\pi}
\sqrt{A^{(\mathrm{fcc})}_+A^{(\mathrm{fcc})}_-}\,~~~~~~~~~~~~{\rm
(quarks)}\ . \ee Using the
 expansion (\ref{dev}) one can get an approximate formula giving
 a power series of $\Delta/qv$.

\section{Specific heat of phonons}
Besides Fermi quasi-particles we consider also the massless
Nambu-Goldstone bosons (NGB). Even though we expect a
parametrically smaller ${\cal O}(T^3)$ contribution in this case,
the contribution of the NGB might be relevant for future
applications, e.g. for the calculation of thermal conductivity.
For two flavor QCD in the crystalline phase the only NGB are
phonons \cite{Casalbuoni:2002pa,Casalbuoni:2002my}.
\subsection{One or two plane waves} Let us begin with the phonon
Lagrangian for the plane wave \be {\cal L}=\frac 1 2\left(
{\dot\phi}^2-v_\parallel^2(\nabla_\parallel\phi)^2-v_\perp^2|\nabla_\perp\phi|^2
\right)\label{lag0} \, ,\ee where $\phi$ is the phonon field,
$\nabla_\parallel={\bf n}\cdot\bm\nabla$,
${\bm\nabla}_\perp={\bm\nabla}-\bf n\nabla_\parallel$ and ${\bf
n}$ along the $z-$axis, i.e. the direction of $\bf q$. In
\cite{Casalbuoni:2002pa} we found \be v_\parallel^2=
\cos^2\theta_q~,~~~~~~~~\ v_\perp^2=\frac 1 2\,\sin^2\theta_q\
,\ee with $\cos\theta_q=\delta\mu_2/q=0.833$. Here $\delta\mu_2$
is the $\delta\mu$ value corresponding to the second order phase
transition from the LOFF phase to the normal one. The dispersion
law, relating the phonon quasi-momentum $\bf k$ and energy
$\omega$, therefore is\be \omega ({\bf k})=\sqrt{
v_\perp^2(k_x^2+k_y^2)+v_\parallel^2k_z^2 }\ .\ee Let $N$ be the
total number of oscillatory modes. There is an oscillator for any
quark pair and the total number of quark pairs is given by the
available phase space, i.e. the pairing region. If $V$ is the
available volume, then \be \frac{ N }V=\frac{g}2
\int\frac{d^3k}{(2\pi)^3} ,\label{37}\ee where $g= 4\times 2$ and
we divide by two due to the fact that there are two quarks in the
pair. The integration over the pairing region gives \be \frac N
V\simeq \frac g 2 \ \frac{4\pi\mu^2\zeta}{(2\pi)^3}\,(2\delta)\,\
. \ee A factor $2\delta$ arises from the integration over the
longitudinal residual momentum $\xi$; $\delta$ is the ultraviolet
cutoff, of the order of $\mu$ while $\zeta$ is the fraction of the
phase space available for pairing in the LOFF phase. It is
estimated of the order of $\Delta/q$.

The ratio $N/V$ can be expressed in terms of a cutoff frequency
by a procedure analogous to the introduction of the Debye
frequency for ordinary crystals. We write\be \frac{ N
}V=\int_0^{\omega_D} f(\omega)d\omega\label{22}\ee where
 $f(\omega)$ is given by
\be f(\omega)=\frac g
2\int\frac{d^3k}{(2\pi)^3}\delta(\omega-\omega({\bf
k}))=\frac{g\omega^2}{4\pi^2v_\perp^2 v_\parallel}
 \, .\ee Substituting in (\ref{37})
we see that $\omega_D$ is of the order of $\mu(\Delta/q)^{1/3}$
 therefore large (excluding the second order phase transition region).
 More precisely we get the formula
  \be \frac{N}{V}=\frac{g\omega_D^3}{12\pi^2v_\perp^2
  v_\parallel}\ .\ee

These results also hold in the case of the strip, i.e.  two
antipodal plane waves. As a matter of fact one can prove,
following the same procedure  as in \cite{Casalbuoni:2002pa},
that the effective Lagrangian for the phonon field is still given
by (\ref{lag0}).

The specific heat per unit volume at small temperatures
($\omega_D\gg T$), is given by ($k_B=\hbar=1$):\be
 c_v=\frac{4 N
\pi^4}{5V}\,\left(\frac{T}{\omega_D}\right)^3\,,\label{cv}\ee and
therefore \be
 c_v^{(\mathrm{FF})}=c_v^{(\mathrm{s})}=\frac{ 8\pi^2
}{15v_\perp^2
  v_\parallel}\,T^3\,~~~~~~{\rm (phonons)}\ .\label{cv4}\ee
  This
result holds for the two considered structures (one or two plane
waves).

 \subsection{Face-centered-cube} For the FCC crystal structure the phonon Lagrangian is
 given by \cite{Casalbuoni:2002hr}\be {\cal L}=\frac 1
2\sum_{i=1,2,3}({\dot\phi}^{(i)})^2-\frac a 2
\sum_{i=1,2,3}|{\bm\nabla}\phi^{(i)}|^2- \frac b 2
\sum_{i=1,2,3}(\de_i\phi^{(i)})^2-
c\sum_{i<j=1,2,3}\de_i\phi^{(i)}\de_j\phi^{(j)}\,.\label{lag}\ee
In \cite{Casalbuoni:2002my} we found \be a=1/12~~~~~~b=0~,~~~~~
c=(3\cos^2\theta_q-1)/12\,.\label{1bis}\ee
 Eq. (\ref{22}) is still valid with\be f(\omega)=\frac 1
3\sum_{r=1}^3\,\frac g
2\int\frac{d^3k}{(2\pi)^3}\delta(\omega-\omega_r({\bf k}))
 \ ,\ee
 since $\omega_r(\vec k)=v_r({\bf\hat n}) k$ we get
 \be f(\omega)=\frac{\omega^2 g}{48\pi^3}
 \sum_{r=1}^3\int \frac{d{\bf\hat n}}{v^3_r({\bf\hat
 n})}\,=\,\frac{\omega^2 g}{24\pi^2}\,K\,.\ee
Here the velocities $v^2_r=v^2_r({\bf\hat n}) $ are the
eigenvalues of the matrix \be  \left( \matrix{ a \,+\,b\,n_1^2 &
c\,n_1 n_2 & c\, n_1 n_3 \cr c\, n_1 n_2 & a\, + b\,n_2^2 & c\,
n_2 n_3\cr c\, n_1 n_3 & c\, n_2 n_3 & a\, +b\,n_3^2 }
\right)\label{matrix}\, , \ee with $n_1=\sin\theta\cos\varphi$,
$n_2=\sin\theta\sin\varphi$, $n_3=\cos\theta$.

 To get the Debye
frequency  and the specific heat we use the numerical result\be
K=\frac 1{2\pi} \sum_{r=1}^3\int \frac{d{\bf\hat
n}}{v^3_r({\bf\hat
 n})}\approx 3.3\times 10^{2}\ee
  corresponding to the values (\ref{1bis}) of the
 parameters. For the specific heat we get, similarly to Eq. (\ref{cv4}), the result
 \be
 c_v^{(\mathrm{fcc})}=\frac{ 4\pi^2
}{15}\,K\,T^3\,~~~~~~~~~~~~{\rm (phonons)}\ .\label{cv4315}\ee One
can note that numerically
$c_v^{(\mathrm{fcc})}>c_v^{(\mathrm{s})}$.

\section{Conclusions and Outlook}

We have considered three different crystalline structures for
high density QCD in the LOFF phase: one plane wave, the strip,
and the favorite face-centered-cubic strucure. Gapless fermionic
quasi-particles are found to exist even in the presence of quark
mass terms.
 These gapless fermions provide for the dominant contributions to the thermal properties
 for low temperatures, such as of possible interest for compact stars. Typically they
 contribute to specific heats with terms linear in $T$ which are dominant at low $T$. The
 massless phonons contribute instead as $T^3$
 while gapped particles in the homogeneous case (2SC) have an exponential
 suppression at low $T$.
The previous calculations could be applied to the evaluation of
transport properties of compact stars. At the present a reliable
model for the LOFF phase in compact stars is still lacking, since
the study of the inhomogeneous superconducting phase of  QCD with
three flavors has not yet been performed. Nevertheless one can
make some conjecture by imagining the existence of the LOFF phase
inside the star, but outside the inner core, where one might
suppose the existence of quark matter in the CFL phase. In this
analysis the thermal properties of the homogeneous and
inhomogeneous phases would be different since for the homogeneous
case the fermion specific heats would be exponentially suppressed,
while for the inhomogeneous case the vanishing with the
temperature would be linear. An application of previous results
might be the evaluation of the thermal conductivity in the LOFF
phase for the existing model of 2-flavor QCD.  We plan to come to
this subject in the future.

\vskip0.5cm\noindent {\bf Acknowledgments}:  We would like  to
thank K. Rajagopal for an enlightening discussion. We wish also to
thank F. Sannino for useful discussions.

\end{document}